\documentclass[preprint,pra,showpacs,nofootinbib]{revtex4}
\def\be{\begin{equation}}
\def\ee{\end{equation}}
\def\ben{\begin{eqnarray}}
\def\een{\end{eqnarray}}

\begin{document}

\begin{flushright}
\vspace*{1cm}
\end{flushright}

\title{\bf Brownian motion of a charged test particle \\ in vacuum between two conducting plates}

\author{Hongwei Yu}
\affiliation{ CCAST(World Lab.),
P. O. Box 8730, Beijing, 100080, P. R. China and Department of
Physics and Institute of  Physics,\\ Hunan Normal University,
Changsha, Hunan 410081, China\footnote{Mailing address}}

\author{Jun Chen }
\affiliation{Department of Physics and Institute of  Physics,\\
Hunan Normal University, Changsha, Hunan 410081, China}

\begin{abstract}

 The Brownian motion of a charged test particle
caused by quantum electromagnetic vacuum fluctuations between two
perfectly conducting plates is examined and the mean squared
fluctuations in the velocity and position of the test particle are
calculated.  Our results show that the Brownian motion in the
direction normal to the plates is reinforced in comparison to that
in the single-plate case. The effective temperature associated
with this normal Brownian motion could be three times as large as
that in the single-plate case. However, the negative dispersions
for the velocity and position in the longitudinal directions,
which could be interpreted as reducing the quantum uncertainties
of the particle, acquire positive corrections due to the presence
of the second plate, and are thus weakened.

\end{abstract}

\maketitle \baselineskip=14pt


\section{Introduction }


Quantum vacuum fluctuations have been subjected to extensive
studies since the emergence of quantum theory which has profoundly
changed our conception of empty space or vacuum. Now we believe
that vacuum is not just a synonym of nothingness. Vacuum
fluctuates all the time and may have rich structures. Although
some physical quantities in quantum field theory, such as energy,
are not well-defined in vacuum and we have to use certain
renormalization scheme to make them finite, {\it changes} in the
vacuum fluctuations, however, usually exhibit normal behavior and
can produce observable effects. The Lamb shift and the Casimir
effect are two examples of this.

Recently, the Brownian motion of a charged particle caused by {\it
changes} in the  electromagnetic vacuum  near a perfectly
reflecting plane boundary has been investigated \cite{YF04} and
the effects have been calculated of the modified electromagnetic
vacuum fluctuations due to the presence of the boundary upon the
motion of a charged test particle . In particular, it has been
shown that the mean squared fluctuations in the normal velocity
and position of the test particle can be associated with an
effective temperature of \be
   T_{eff}={\alpha\over \pi}\;{1\over k_B mz^2}=
  1.7\times 10^{-6}\left({{1 \mu m}\over z}\right)^2\;K
= 1.7\times 10^2\left({1\mathring{A}\over z}\right)^2\;K\; ,
\label{eq:effT}
 \ee
where $k_B$ is Boltzmann's constant and $z$ is the distance from
the boundary. This result seems to be experimentally accessible.

However, in addition to the fluctuating quantum field-theoretic
force, a charged particle also feels a classical image charge
force in  the single-plate geometry. This force tends to pull the
particle toward the plate and thus makes it difficult to observe
the random motion of the particle caused by quantum vacuum
fluctuations studied in Ref.~\cite{YF04}.  One way to minimize the
influence of the image charge is to add another parallel plate.
Then, a charged particle moving exactly at midway between the
plates feels no net classical force. Note, however, that this
trajectory is unstable. It can be shown that adding a second plate
could increase the characteristic falling time of the particle to
a plate by approximately an order of magnitude, making it
experimentally more feasible to observe the effects of the quantum
vacuum fluctuations. So, we wish to examine, in the present paper,
the Brownian motion of a charged test particle coupled to the
electromagnetic vacuum fluctuations between two conducting plates.

Finally, it is worth noting that the present problem bears some
analogy to the problem of lightcone fluctuations, where photons
undergo Brownian motion due to modified quantum fluctuations of
the quantized gravitational field~\cite{F95,YF99,YF00,YW}.


\section{Vacuum fluctuation and the Brownian motion of the test particle}


Let us consider the motion of a charged test particle subject to
quantum electromagnetic vacuum fluctuations in the vacuum between
two conducting plates.
 In the limit of small velocities and
assuming that the particle is initially at rest and has a charge
to mass ratio of $ e/m$,  the mean squared speed in the
$i$-direction can be written as (no sum on $i$)
\begin{eqnarray}
\langle{\Delta v_i^2}\rangle&=&{e^2\over
m^2}\;\int_0^t\;\int_0^t\; \langle{E}_i({\mathbf
x},t_1)\;{E}_i({\mathbf x},t_2)\rangle_R\,dt\nonumber\\
&=&{4\pi\alpha\over m^2}\;\int_0^t\;\int_0^t\;
\langle{E}_i({\mathbf x},t_1)\;{E}_i({\mathbf x},t_2)\rangle_R\,dt
 \; ,   \label{eq:lang2}
\end{eqnarray}
where $\alpha$is the fine-structure constant and $
\langle{E}_i({\mathbf x},t_1)\;{E}_i({\mathbf x},t_2)\rangle_R $
is the renormalized electric field two-point function obtained by
subtracting the boundary-independent Minkowski vacuum term.  We
have, for simplicity, assumed that the distance does not change
significantly in a time $t$, so that it can be treated
approximately as a constant. If there is a classical,
nonfluctuating field in addition to the fluctuating quantum field,
then Eq.~(\ref{eq:lang2}) describes the velocity fluctuations
around the mean trajectory caused by the classical field.

Let us now calculate the vacuum expectation of the squared
electric field in the case of two parallel conducting plates. The
two point function for the photon field may be expressed as
\begin{equation}
\label{eq:TwoPoint}
 D^{\mu \nu} (x,x') = \langle 0|  A^\mu (x)\, A^\nu (x') |0
\rangle ={D^{\mu \nu}_{0}} (x-x') +{D^{\mu \nu}_{R}} (x,x')\;,
\end{equation}
where $D_{0}^{\mu\nu}(x-x')$ is the two point function in the
usual Minkowski vacuum, and the renormalized two point function,
${D^{\mu \nu}_{R}} (x,x')$, is the correction induced by the
presence of boundaries which can be obtained by the  method of
images~\cite{BM69}. In the Feynman gauge, we have, assuming one
plate is at $z=0$ and the other at $z=a$,
\begin{equation}
\label{eq:TwoPoint0}
D_{0}^{\mu\nu}(x-x')=\frac{\eta^{\mu\nu}}{4\pi^{2}(\Delta
t^{2}-\Delta \mathbf{x}^{2})}
\end{equation}
and
\begin{eqnarray}
 \label{eq:rgf2}
D_{R}^{\mu\nu}(x,x')=-\sum^{\infty}_{n=-\infty}\frac{\eta^{\mu\nu}+2n^{\mu}n^{\nu}}
{4\pi^{2}(\Delta t^{2}-\Delta x^{2}-\Delta y^{2}-(z+z'+2na) ^{2})}
\nonumber\\+ \sum^{\infty}_{n=-\infty}{'} \frac{\eta^{\mu\nu}}
{4\pi^{2}(\Delta t^{2}-\Delta x^{2}-\Delta y^{2}-(z-z'+2na)
^{2})}\label{eq3}\; ,
\end{eqnarray}
Here and after a prime means that the $n=0$ term is omitted in the
summation, $\eta^{\mu\nu}$= diag(1,-1,-1,-1) and the unit normal
vector $n^{\mu}= (0,0,0,1)$. Note that the two-point function
Eq.~(\ref{eq:rgf2}) is constructed in such a way that the
tangential components of the electric field two-point function
vanish on the conducting plates \cite{BM69}, i.e., whenever $z= 0$
or $z =a$.  At a point a distance $z$ from the plate at $z=0$, the
components of the renormalized electric field two-point function,
$\langle{\mathbf E}({\mathbf x},t_1)\;{\mathbf E}({\mathbf
x},t_2)\rangle_R$, are\footnote{Lorentz-Heaviside units with
$c=\hbar =1$ will be used here.}
\begin{eqnarray}
&&\langle E_{x}({\mathbf{x}},t')E_{x}({\mathbf{x}},t'')\rangle_R
=\langle E_{y}({\mathbf{x}},t')E_{y}({\mathbf{x}},t'')\rangle_R
\nonumber\\&&\quad
=\frac{1}{\pi^{2}}\Bigg\{\sum^{\infty}_{n=-\infty}{'} \frac{\Delta
t^2+4n^2a^2} {(\Delta
t^2-4n^2a^2)^3}-\sum^{\infty}_{n=-\infty}\frac{\Delta
t^2+4(na+z)^2} {[\Delta t^2-4(na+z)^2]^3}\Bigg\}\;, \label{eq:Ex}
\end{eqnarray}
and
 \begin{eqnarray}
 \langle E_{z}({\mathbf{x}},t')E_{z}({\mathbf{x}},t'')\rangle_R
 =\frac{1}{\pi^{2}}\Bigg\{\sum^{\infty}_{n=-\infty}{'}
 \frac{1}{(\Delta t^2-4n^2a^2)^2}+
 \sum^{\infty}_{n=-\infty}\frac{1}{[\Delta
 t^2-4(na+z)^2]^2}\Bigg\}\;.\nonumber\\
 \label{eq:Ez}
 \end{eqnarray}
Substituting the above results into Eq.~(\ref{eq:lang2}) and
carrying out the integration, we find that the velocity
dispersions in the parallel directions are given by \ben
\label{eq:vx}
 \langle \Delta v_x^2\rangle=\langle \Delta v_y^2\rangle&=&{e^2\over
m^2}\;\int_0^t\;\int_0^t\;\langle{E}_x({\mathbf x},t')\;{E}_x({\mathbf
 x},t'')\rangle_R\; dt'\;dt''\nonumber\\
 &=& {e^2\over \pi^2 m^2}\;\Bigg\{\sum^{\infty}_{n=-\infty}{'} f_L(na,t)-\sum^{\infty}_{n=-\infty} f_L(na+z,t)\Bigg\}\;.
\een
 Here we have defined
 \be
 f_L(x,t)\equiv\frac{t^2}{8x^2(t^2-4x^2)}-\frac{t}{64|x|^3}\ln\left(\frac{t+2|x|}{t-2|x|}\right)^2
 \label{eq:Fxt}\;.
\ee
 The mean squared position fluctuations  can
be calculated as follows
 \ben
 \langle\Delta x^2\rangle=\langle\Delta y^2\rangle&=&
\int_0^t\;dt_1\;\int_0^{t_1}\;dt'\;\int_0^t\;dt_2
\;\int_0^{t_2}\;dt''\; \langle{E}_x({\mathbf
x},t')\;{E}_x({\mathbf
 x},t'')\rangle_R\nonumber\\
&=& {e^2\over \pi^2 m^2}\;\Bigg\{\sum^{\infty}_{n=-\infty}{'}
g_L(na,t)-\sum^{\infty}_{n=-\infty} g_L(na+z,t)\Bigg\}\;,
 \een
  with
 \be
 g_L(x,t)\equiv\frac{e^2}{\pi^2m^2}\, \left[-{t^3\over192|x|^3}\,
\ln\left(\frac{t+2|x|}{t-2|x|}\right)^2 +{t^2\over 24x^2}+{1\over
6}\ln\left(\frac{t^2-4x^2}{4x^2}\right) \right]\;.
 \ee
In the transverse direction, the velocity dispersion is
 \ben
 \langle \Delta v_z^2\rangle&=&{e^2\over \pi^2 m^2}\;\int_0^t\;\int_0^t\;
 \langle{E}_z({\mathbf x},t')\;{E}_z({\mathbf
 x},t'')\rangle_R\;dt'\;dt''\nonumber\\
&=&{e^2\over \pi^2 m^2}\;\Bigg\{\sum^{\infty}_{n=-\infty}{'}
f_T(na,t)+\sum^{\infty}_{n=-\infty} f_T(na+z,t)\Bigg\}\;,
 \een
where $f_T(x,t)$ is defined as
 \be
f_T(x,t)={t\over 32|x|^3}\ln\biggl( {2|x|+t\over
 2|x|-t}\biggr)^2\;.
 \ee
The corresponding position fluctuation is
 \ben
 \langle \Delta z^2\rangle
&=& \int_0^t\;dt_1\;\int_0^{t_1}\;dt'\;\int_0^t\;dt_2
\;\int_0^{t_2}\;dt''\; \langle{E}_z({\mathbf
x},t')\;{E}_z({\mathbf
 x},t'')\rangle_R\nonumber\\
&=& {e^2\over \pi^2 m^2}\;\Bigg\{\sum^{\infty}_{n=-\infty}{'}
g_L(na,t)-\sum^{\infty}_{n=-\infty} g_L(na+z,t)\Bigg\}\;,
 \een
  with
 \be
 g_L(x,t) = \frac{e^2}{\pi^2 m^2}\, \left[\frac{t^2}{24 x^2}
+ \frac{t^3}{96 |x|^3}\, \ln\left(\frac{t+2|x|}{t-2|x|}\right)^2 +
\frac{1}{6}\, \ln\left(\frac{t^2-4x^2}{4x^2}\right)\right]\; .
 \label{eq:z2}
 \ee
 It does not seem to be an easy task to get a closed-form result
for the above summations. Now we wish to discuss two special cases
of interest, i.e., the cases of $a\gg t$ and $a\ll t$.


\subsection{The $a\gg t$ case}


In the $a\gg t$ case,  the mean squared fluctuations in the
velocity of the particle can be approximately evaluated as follows
 \begin{eqnarray}
\langle \Delta v_x^2\rangle=\langle \Delta v_y^2\rangle &\approx&
\frac{e^2}{\pi^2 m^2} \Bigg[-\frac{\pi^4t^2}{720a^4}
+\frac{\pi^4t^2(2+\cos\frac{2\pi
z}{a})\csc^{4}\frac{z\pi}{a}}{48a^4}
\nonumber\\&&\quad\quad\quad-\frac{t^2}{16(a-z)^4}-\frac{t^2}{16z^4}
-f_L(z,t)-f_L(a-z,t)\; \Bigg]\;,
 \end{eqnarray}
and
\begin{eqnarray}
\langle \Delta v_z^2\rangle&\approx& \frac{e^2}{\pi^2 m^2}
\Bigg[\frac{\pi^4t^2}{720a^4} +\frac{\pi^4t^2(2+\cos\frac{2\pi
z}{a})\csc^{4}\frac{z\pi}{a}}{48a^4}
\nonumber\\&&\quad\quad\quad-\frac{t^2}{16(a-z)^4}-\frac{t^2}{16z^4}
+f_T(z,t)+f_T(a-z,t)\; \Bigg]\;,
\end{eqnarray}
 where we have used
 \be
\sum^{\infty}_{n=-\infty} \frac{1}{(na+z)^4} =
\frac{\pi^4(2+\cos\frac{2\pi
z}{a})\csc^{4}\frac{z\pi}{a}}{3a^4}\;.
 \ee
It should be pointed out that the above expressions are singular
at  $t=2z$ or $t=2(a-z)$. This corresponds to a time interval
equal to the round-trip light travel time between the particle and
 the plate at $z=0$ or $z=a$.  Presumably, this might be a result of our
assumption  of a rigid perfectly reflecting plane boundary,  and
would thus be  smeared out in a more realistic treatment,  where
fluctuations in the position of  the plates are taken into
account. Note that the above results are  symmetric
 under $z\leftrightarrow a-z $ as they should be by the symmetry of the
 system and furthermore  $\langle \Delta v_x^2\rangle$ is regular
 in limits of both $z \rightarrow 0$ and $z \rightarrow a $, which
 can be seen by expanding $\langle \Delta v_x^2\rangle$ around $z=0$.
 This may come as a surprise at the first glance. It can, however, be
 understood as a result of the fact that the tangential components
 of the electric field two-point function vanish at the conducting boundaries.
 $\langle \Delta v_z^2 \rangle$, on the other hand, diverges as the boundaries are approached
 and it is a reflection of the divergence of the normal electric field
two-point function at the boundaries. If we further assume that
$a\gg z$, then
 \be
 \label{eq:Vx1}
\langle \Delta v_x^2\rangle=\langle \Delta v_y^2\rangle \approx
\frac{e^2}{\pi^2 m^2}\bigg[- f_L(z,t)+ {1\over8}\bigg( {t\over
a}\bigg)^4\;\bigg( {z\over a}\bigg)^8\; {1\over t^2}\;\bigg ]\;,
 \ee
 and
 \be
 \label{eq:Vz2}
\langle \Delta v_z^2\rangle\approx \frac{e^2}{\pi^2 m^2}\bigg[
f_T(z,t)+ {\pi^4t^2\over 360a^4}\;\bigg ]\;.
 \ee
A comparison of the above results with Eqs.~(9,11) of
Ref.\cite{YF04} shows that the two last terms in the above
equations are the corrections introduced by the presence of the
second plate which, in the approximation being considered, can be
regarded as being very far away from the other plate.
 For $ t\gg z$, Eq.~(\ref{eq:Vx1}) and  Eq.~(\ref{eq:Vz2}) become
 \be
 \langle \Delta v_x^2\rangle=\langle \Delta v_y^2\rangle
\approx\;-{e^2\over 3\pi^2 m^2}{1\over t^2}+{e^2\over 8\pi^2
m^2}\bigg( {t\over a}\bigg)^4\;\bigg( {z\over a}\bigg)^8\; {1\over
t^2}\;,
 \ee
and \be
 \langle \Delta v_z^2\rangle\approx\;
 {e^2\over 4 \pi^2 m^2}{1\over
z^2} +{\pi^2e^2\over 360m^2a^2}\bigg({t\over a}\bigg)^2+ {e^2\over
3\pi^2 m^2}{1\over t^2} \;. \label{eq:vz2}
 \ee
 The correction term has an opposite sign in the longitudinal directions and it makes the
 dispersion less negative than what it would be with just a single
 plate. While the dispersion in the transverse direction acquires an
 positive correction and is thus greater than that in the
 single-plate case.

Similarly, we have for the dispersions in position
 \ben
 \langle
\Delta x^2\rangle=\langle \Delta y^2\rangle
&\approx&\frac{e^2}{\pi^2 m^2} \Bigg[-\frac{\pi^4t^4}{2880a^4}
+\frac{\pi^4t^4(2+\cos\frac{2\pi
z}{a})\csc^{4}\frac{z\pi}{a}}{192a^4}
\nonumber\\&&\quad\quad\quad-\frac{t^4}{64(a-z)^4}-\frac{t^4}{64z^4}
-g_L(z,t)-g_L(a-z,t)\; \Bigg]\;,
 \een
 and
\ben
 \langle \Delta z^2\rangle &\approx&\frac{e^2}{\pi^2 m^2}
\Bigg[\frac{\pi^4t^4}{2880a^4} +\frac{\pi^4t^4(2+\cos\frac{2\pi
z}{a})\csc^{4}\frac{z\pi}{a}}{192a^4}
\nonumber\\&&\quad\quad\quad-\frac{t^4}{64(a-z)^4}-\frac{t^4}{64z^4}
+g_L(z,t)+g_L(a-z,t)\; \Bigg]\;.
 \een
Their limiting forms for $a\gg z$ and $ t\gg z$ are
 \be
 \langle \Delta x^2\rangle= \langle \Delta y^2\rangle\approx\;-{e^2\over
3\pi^2m^2}\ln(t/2z)+ \frac{e^2}{32\pi^2 m^2}\bigg( {t\over
a}\bigg)^4\;\bigg( {z\over a}\bigg)^8\;,
 \ee
and
 \be
 \langle \Delta z^2\rangle \approx \frac{e^2}{\pi^2 m^2}\, \left[
\frac{t^2}{8 z^2} + \frac{1}{3}
\ln\left(\frac{t}{2z}\right)+{\pi^4\over 1440}\bigg({t\over
a}\bigg)^4  \right] \;. \label{eq:z2_large}
 \ee


\subsection{The $t\gg a$ case}


 We now turn to the $t\gg a$ case, which is of more significance as far as the experimental measurement of the effects
 are concerned.  Let us first write the
velocity dispersions in the parallel directions as
 \be
 \langle \Delta v_x^2\rangle=\langle \Delta v_y^2\rangle =
 \frac{e^2}{\pi^2m^2}\Big[-f_L(z,t)-f_L(z-a,t)+h(z,t)\Big]\;,
 \label{eq:Vxta}
 \ee
 where
 \begin{eqnarray}
h(z,t)\equiv\sum^{+\infty}_{n=1}\Big[2f_L(na,t)-f_L(z+na,t)\Big]
-\sum^{+\infty}_{n=2}f_L(z-na,t)\;.
\end{eqnarray}
Now the summation can be estimated by integration. Defining
$\gamma=t/2a$,  we have, for example,
\begin{eqnarray}
\sum^{+\infty}_{n=1}2f_L(2na,t)&\approx &
\frac{\gamma}{t^2}\int^{\infty}_{1/\gamma}\frac{1}{x^2(1-x^2)}dx
-\frac{\gamma}{2t^2}\int^{\infty}_{1/\gamma}\frac{1}{x^3}\ln\Big(\frac{1+x}{1-x}\Big)dx
\nonumber\\&&\equiv\gamma w(1/\gamma)\;.
\end{eqnarray}
It then follows that
\begin{eqnarray}
h(z,t)\approx\gamma\Big[w(1/\gamma)-\frac{1}{2}w(1/\gamma+2z/t)-
\frac{1}{2}w(2/\gamma-2z/t)\Big]\;.
\end{eqnarray}
Performing the integration and expanding the result as a power
series of $1/\gamma$, we obtain
 \be
 \langle \Delta
 v_x^2\rangle=\langle \Delta v_y^2\rangle \approx
 \frac{e^2}{\pi^2m^2}\Bigg[-\frac{1}{3t^2}+\frac{32a^2-24z(a-z)}{15t^4}\Bigg]\;.
 \ee
Similarly, we find
\be
 \langle \Delta v^{2}_{z}\rangle\approx
\frac{e^2}{\pi^2m^2}
\Bigg[\frac{a^2-2z(a-z)}{4z^2(a-z)^2}+\frac{1}{2a^2}+\frac{3}{4(a+z)(2a-z)}
-\frac{1}{t^2}\Bigg]\;.
 \ee
 The above results have been written in such a way that the
 symmetry under $z \leftrightarrow a-z$ is manifest.
 In the further limit $a\gg z$, these velocity dispersions become
 \be
 \langle \Delta
 v_x^2\rangle=\langle \Delta v_y^2\rangle \approx
 \frac{e^2}{\pi^2m^2}\Bigg[-\frac{1}{3t^2}+\frac{32a^2}{15t^4}\Bigg]\;,
 \ee
and
\begin{eqnarray}
\langle \Delta v^{2}_{z}\rangle\approx\frac{e^2}{\pi^2m^2}
\Bigg(\frac{1}{4z^2}+\frac{7}{8a^2} -\frac{1}{t^2}\Bigg)\;.
\end{eqnarray}
It is interesting to note
 that the mean squared velocity fluctuation in the
 directions parallel to the plates dies off quickly as time
 progresses
 and it is basically a transient effect.
Unlike the velocity dispersion in the parallel directions, that in
the direction perpendicular to the plate approaches a nonzero
constant value at late times. It is interesting to note that here
no dissipation is needed for $\langle \Delta v_i^2\rangle$ to be
bounded at late times in contrast to the Brownian motion due to
thermal noise.

The mean squared fluctuations in position in both parallel and
perpendicular directions can also be found in the same way as
follows
 \ben
 \langle\Delta x^2\rangle\approx
\frac{e^2}{\pi^2m^2}&&\Bigg[-\frac{1}{3}\ln\frac{t}{2z}-\frac{1}{3}\ln\frac{t}{2(a-z)}
-\frac{2}{3}\ln\frac{t}{2a}+\frac{a+z}{3a}\ln\frac{t}{2(a+z)}
\nonumber\\&&\quad+\frac{2a-z}{3a}\ln\frac{t}{2(2a-z)}\Bigg]\;,
\end{eqnarray}
and
\begin{eqnarray}
\langle\Delta z^2\rangle &\approx& \frac{e^2}{\pi^2m^2}\Bigg[
\frac{t^2(a^2-2z(a-z))}{8z^2(a-z)^2}+\frac{t^2}{4a^2}
+\frac{3t^2}{8(a+z)(2a-z)}+\frac{1}{3}ln\frac{t^2}{4z(a-z)}
\nonumber\\&&\quad-\frac{2}{3}ln\frac{t}{2a}
-\frac{a+z}{3a}ln\frac{t}{2(a+z)}
-\frac{2a-z}{3a}ln\frac{t}{2(2a-z)}\Bigg].
\end{eqnarray}
Their limiting forms for $a\gg z$ are
\begin{eqnarray}
\langle\Delta x^2\rangle\approx\frac{e^2}{\pi^2m^2}
\Bigg(-\frac{1}{3}ln\frac{t}{2z}-\frac{2}{3}ln2\Bigg)\;,
\end{eqnarray}
and
\begin{eqnarray}
\langle\Delta z^2\rangle\approx\frac{e^2}{\pi^2m^2}\Bigg(
\frac{t^2}{8z^2}+\frac{7t^2}{16a^2}+\frac{1}{3}ln\frac{t}{2z}-\frac{2}{3}ln\frac{t}{2a}
-\frac{2}{3}ln\frac{t}{4a}\Bigg)\;.
\end{eqnarray}
Finally, let us note that all dispersions attain their extremal
values at $z=a/2$ as listed below
 \begin{eqnarray}
\langle \Delta v^{2}_{x}\rangle\approx
\frac{e^2}{\pi^2m^2}\Bigg[-\frac{1}{3t^2}+\frac{26a^2}{15t^4}\Bigg]\label{eq:vxma}\;,
 \end{eqnarray}
 \begin{eqnarray}
\langle \Delta v^{2}_{z}\rangle\approx
\frac{e^2}{\pi^2m^2}\Bigg(\frac{17}{6a^2}-\frac{1}{t^2}\Bigg)\label{eq:vzma}\;,
\end{eqnarray}
 \begin{eqnarray}
\langle\Delta x^2\rangle\approx \frac{e^2}{\pi^2m^2}
\Bigg(-\frac{2}{3}ln\frac{t^2}{2a^2}+ln\frac{t}{3a}\Bigg)\label{eq:xma}\;,
 \end{eqnarray}
 \begin{eqnarray}
\langle\Delta z^2\rangle\approx
\frac{e^2}{\pi^2m^2}\Bigg(\frac{17t^2}{12a^2}+\frac{2}{3}ln
2-ln\frac{t}{3a}\Bigg)\label{zma}\;.
 \end{eqnarray}

 If we compare the above results with those in the
case of a single plate when $z=a/2$ \cite{YF04}, we can see that
the dispersions in the normal direction to the plates get
amplified roughly by a factor of $17/6\approx 2.8$, i.e., the
dispersions in the present case are approximately three times that
of the single-plate case.

One may wonder if the $t\gg a$ case is meaningful, since the
particle will ultimately fall to one of the plates due to the
classical image force. There is no exception to this destiny even
if it is placed at the midway, when quantum fluctuations are
considered. It can be shown that characteristic falling time,
$t_c$, is given approximately  by $\sqrt{\frac{mz_{0}^3}{e^2}}$.
Here $z_0$ is the initial  distance to a plate.  For the
calculations in the $t\gg a$ case to make sense, we must have
$t_c\gg
 a $. Taking an electron initial at the midway as an example, this leads to the  constraint on  $
 a\gg 2.4\times 10^{-10}cm = 2.4\times 10^{-4}\mu m\;.
 $ Another issue is the condition for our assumption in
Eq.~(\ref{eq:lang2}) to be valid that the particle does not
significantly change its position, i.e., $\langle \Delta
z^2\rangle \ll z$. It can be shown that this requirement yields
$t\ll (mz)z$.  For an electron initially at $z=a/2$ at $t=0$, this
means $t\ll 10^{10}(a/1cm)a$. So, there is much room for both
$t\gg a$ and our assumption to be fulfilled.

\section{Summary}

In summary, we have examined the Brownian motion of a charged test
particle caused by quantum electromagnetic vacuum fluctuations
between two perfectly conducting plates and calculated the mean
squared fluctuations in the velocity and position of the test
particle. Our results show that the Brownian motion in the
transverse direction to the plates is reinforced in comparison to
that in the single-plate case. The effective temperature
associated with this transverse Brownian motion could be three
times as large as that in the case of the single-plate. And this,
together with the fact that adding the second plate could increase
the characteristic falling time of the particle by roughly an
order of magnitude, suggests that the Brownian motion caused by
the vacuum fluctuations would be easier to be observed in the
two-plate case. Our calculations also reveal that the negative
dispersions for the velocity and position in the longitudinal
directions, which could be interpreted as reducing the quantum
uncertainties of the particle, acquire positive corrections due to
the presence of the second plate, and are hence weakened.

\begin{acknowledgments}
 This work was supported in part  by the National Natural Science
Foundation of China  under Grant No. 10375023, the National Basic
Research Program of China under Grant No. 2003CB71630.
\end{acknowledgments}

\end{document}